\documentclass[12pt, letterpaper]{article}
\usepackage[top=1in, bottom=1.5in, left=1in, right=1in]{geometry}
\usepackage[utf8]{inputenc}
\usepackage{booktabs}
\usepackage{hyperref}
\usepackage{graphicx}
\usepackage{amssymb}
\usepackage{multicol}
\title{unVEil the darknesS of The gAlactic buLgE (VESTALE)}
\date{} 

\begin{document}
\maketitle
\vspace{-1.5cm}
 \noindent G. Bono,$^{1,2}$ M. Dall'Ora,$^{3}$ M. Fabrizio,$^{2,4}$ J. Crestani,$^{1,2,5}$ V.F. Braga,$^{6,7}$ G. Fiorentino,$^{8}$ G. Altavilla,$^{2,4}$ M.T. Botticella,$^{3}$ A. Calamida,$^{9}$ M. Castellani,$^{2}$ M. Catelan,$^{10}$ B. Chaboyer,$^{11}$ C. Chiappini, $^{37}$ W. Clarkson,$^{12}$ R. Contreras Ramos,$^{6,10}$ O. Creevey,$^{13}$ R. da Silva,$^{2,4}$ V. Debattista,$^{14}$ S. Degl’Innocenti,$^{15,16}$ I. Ferraro,$^{2}$ C.K. Gilligan,$^{11}$ O. Gonzalez,$^{17}$ K. Hambleton,$^{40}$ G. Iannicola,$^{2}$ L. Inno,$^{18}$ A. Kunder,$^{19}$ B. Lemasle,$^{20}$ L. Magrini,$^{18}$ D. Magurno,$^{1,2}$ M. Marconi,$^{3}$ M. Marengo,$^{21}$ S. Marinoni,$^{2,4}$ P.M. Marrese,$^{2,4}$ C.E. Mart\'inez-Vazquez,$^{22}$ N. Matsunaga,$^{23}$ M. Monelli,$^{24,25}$ P.G. Prada Moroni,$^{15,16}$ I. Musella,$^{3}$ M.G. Navarro,$^{26,7,6}$ J. Neeley,$^{27}$ M. Nonino,$^{28}$ A. Pietrinferni,$^{29}$ L. Pulone,$^{2}$ M.R. Rich,$^{38}$ V. Ripepi,$^{3}$ G. Sacco,$^{18}$ A. Saha,$^{22}$ M. Salaris,$^{30}$ C. Sneden,$^{31}$ P.B. Stetson,$^{32,33}$ R.A Street, $^{39}$ R. Szabo,$^{34}$ M. Tantalo,$^{1}$ E. Tognelli,$^{35,15,16}$ M. Torelli,$^{2}$ E. Valenti,$^{36}$ A.R. Walker,$^{22}$ and M. Zoccali$^{10,6}$, \\
\centerline{\textit{with the support of the LSST Transient and Variable Stars Collaboration}.}

\vspace{0.5cm}
\begin{center}
November 29, 2018
\end{center} 
\vspace{0.5cm}

\begin{abstract}
The main aim of this experiment is to provide a complete census of
old (t $\ge 10$ Gyr, RR Lyrae, type II Cepheids, red horizontal branch), intermediate
age (red clump, Miras) and young (classical Cepheids) stellar
tracers across the Galactic Bulge. To fully exploit the unique
photometric quality of LSST images, we plan to perform a {\em Shallow
minisurvey} ($ugrizy$, -20$\lesssim$l$\lesssim$20 deg, 
-15$\lesssim$b$\lesssim$10 deg)
and a {\em Deep minisurvey} ($izy$, -20$\lesssim$l$\lesssim$20 deg, 
-3$\lesssim$b$\lesssim$3 deg).
The former one is aimed at constraining the 3D structure of the galactic 
Bulge across the four quadrants, and in particular, the transition 
between inner and outer Bulge. The $u,g,r,i,z,y$ LSST bands provide fundamental
diagnostics to constrain the evolutionary properties of low and
intermediate-mass stars when moving from a metal-poor to a metal-rich
regime.  The deep minisurvey is aimed at tracing RR Lyrae, Red Clump 
stars, Miras and classical Cepheids in highly reddened regions of 
the Galactic center.
These images will allow us to investigate the role that baryonic mass 
and dark matter played in the early formation and evolution of the MW.     

\end{abstract}


\section{White Paper Information}
\begin{enumerate} 
\item {\bf Science Category:} the basic science theme of this project is the Milky Way Structure and Formation. However, since it is based on variable stars as population tracers and distance indicators, it is also related to the Explore the Changing Sky theme.
\item {\bf Survey Type Category:} mini survey. 
\item {\bf Observing Strategy Category:} this is a project aimed at detecting variable stars in the MW Bulge. Therefore, it is an integrated program with science that hinges on the combination of pointing and detailed observing strategy. 
\item {\bf Author Information}\\
$^{1}$Universit\`a di Roma Tor Vergata \\
$^{2}$INAF--Osservatorio Astronomico di Roma \\
$^{3}$INAF--Osservatorio Astronomico di Capodimonte \\
$^{4}$Space Science Data Center–ASI \\
$^{5}$Universidade Federal do Rio Grande do Sul \\
$^{6}$Instituto Milenio de Astrof\'isica \\
$^{7}$Universidad Andr\'es Bello \\
$^{8}$INAF--OAS Osservatorio di Astrofisica \& Scienza dello Spazio di Bologna \\
$^{9}$Space Telescope Science Institute \\
$^{10}$Pontificia Universidad Cat\'olica de Chile \\
$^{11}$Dartmouth College \\
$^{12}$University of Michigan-Dearborn \\
$^{13}$Universit\`e C\^ote d'Azur \\
$^{14}$University of Central Lancashire  \\
$^{15}$Universit\`a di Pisa \\
$^{16}$INFN, Sezione di Pisa \\
$^{17}$UK Astronomy Technology Centre, Royal Observatory \\
$^{18}$INAF--Osservatorio Astrofisico di Arcetri \\ 
$^{19}$Saint Martin's University \\
$^{20}$Zentrum f\"ur Astronomie der Universit\"at Heidelberg  \\
$^{21}$Iowa State University \\
$^{22}$National Optical Astronomy Observatory \\
$^{23}$The University of Tokyo \\
$^{24}$Instituto de Astrof\'sica de Canarias \\
$^{25}$Universidad de La Laguna \\
$^{26}$Universit\`a di Roma La Sapienza  \\
$^{27}$Florida Atlantic University \\ 
$^{28}$INAF--Osservatorio Astronomico di Trieste  \\
$^{29}$INAF--Osservatorio Astronomico d'Abruzzo \\
$^{30}$Liverpool John Moores University \\
$^{31}$University of Texas \\
$^{32}$Dominion Astrophysical Observatory  \\
$^{33}$National Research Council of Canada \\
$^{34}$Konkoly Observatory \\
$^{35}$INAF--Osservatorio Astronomico di Collurania \\
$^{36}$European Southern Observatory \\
$^{37}$Leibniz Institut fuer Astrophysik Potsdam - AIP\\
$^{38}$Department of Physics \& Astronomy, The University of California\\
$^{39}$Las Cumbres Observatory\\
$^{40}$Villanova University, Dept. of Astrophysics and Planetary Science


\end{enumerate}

\clearpage


\section{Scientific Motivation}
This experiment is aimed at disentangling the stellar content 
of the Galactic Bulge using variable stars, since they have the 
advantage to provide individual distance, age, metallicity and 
reddening estimates. We focus our attention on variables tracing old 
(RR Lyraes, RRLs; Type II Cepheids, TIICs; t$>$10 Gyr), 
intermediate-age (Miras, t$\sim$0.5–10 Gyr), and young 
(Classical Cepheids, CC; t$\sim$10–300 Myr) stellar 
populations (Bono et al. 2015; Matsunaga et al. 2016).
The Galactic Bulge, which is mainly old with a younger tail, 
makes up the 25\% of the total MW stellar mass (Valenti et al. 2016). 
Recent photometric and spectroscopic investigations revealed that 
the Bulge contains two main components. The old and/or metal-poor one, 
traced either with RRL 
or with metal-poor Red Clump (RC) stars, is rounder, rotates slower and 
has a shallower gradient in radial velocity dispersion. The metal-rich 
one is traced with RC stars, it is arranged in a bar that flares up into a 
boxy/peanut structure in its outer region, rotates faster, and
has a steeper gradient in radial velocity dispersion (Ness et al. 2012; 
Rojas-Arriagada et al. 2014; Pietrukowicz et al.
2015; Kunder et al. 2016; Zoccali et al. 2017).
Recent spectroscopic surveys mainly based on either giants 
(BRAVA; Shen et al. 2010) or RC stars (ARGOS; GIBS;
Freeman et al. 2013; Zoccali et al. 2014) 
suggest that Bulge stars undergo cylindrical rotation. On
the other hand, BRAVA-RR used RRLs and found much slower rotations, 
and higher velocity dispersions (Barbuy et al. 2018). Moreover, it is 
not clear yet whether Bulge RRLs trace either the main Bar or the spheroidal
component (D\'ek\'any et al. 2013; 
Pietrukowicz et al. 2015; Kunder et al. 2018).

\vspace{-0.35truecm}
\begin{center}
\bf Why a shallow minisurvey
\end{center}
\vspace{-0.35truecm}

$\bullet$ {\em 3D Bulge Structure} -- 
We have recently developed a new algorithm to estimate reddening distance and metallicity
(REDIME) by using optical/NIR (BVIJHK) bands (Bono et al. 2018). The key advantage of
this approach is that we can provide the 3D structure of the Bulge, a 3D reddening map and 
a homogeneous metallicity distribution of the entire sample of RRLs using "blue" ($ugr$) and 
"red" ($izy$) LSST bands. The zero-point of the metallicity distribution might be affected
by the accuracy of the adopted reddening law and of the distance diagnostic. However, 
we are interested in the differential variation and the accuracy is $\sim$0.2 dex. 
This means new constraints on the occurrence of a metallicity gradient across the 
Bulge (Hill et al. 2011; Zoccali et al. 2017); the shape 
of the Bulge in the four quadrants; the real extent and geometry of both inner and 
long Bar (Hammersley et al. 2000; Athanassoula 2005). We are also interested in 
estimating the position angle and the inclination of the Bar by using old (RRLs, TIICs), 
intermediate age (RC, Miras) and young (classical Cepheids) stars to constrain its 
secular stability (Wegg \& Gerhard 2013).

$\bullet$ {\em Bulge stellar populations} -- 
The current structure of the Bulge mainly relies on RC stars, i.e. old/intermediate 
age stellar tracers. However, solid theoretical (Salaris et al. 2003) and empirical
(Stetson et al. 2011) evidence indicates that RC stars are intermediate-mass, central helium burning
stars, while red HB stars are low-mass, central helium burning stars. The difference in visual magnitude, at fixed metal
content, is at least of the order of 0.5 magnitude, while the optical colors are, at solar chemical compositions, quite
similar (see Fig.~1). The two subpopulations have never been identified in the Bulge due to a mix between photometric
error and differential reddening. Data plotted in Fig.~2 indicate that LSST can trace the variation 
between the two different sub-populations across the entire Bulge. 
We also plan to use the equivalent of the C$_{UBI}$ ([U-B]-[B-I]) photometric index (Monelli et al. 2016), 
but for the SDSS bands C$_{ugi}$  ([u-g]-[g-i]) to separate old and intermediate-age Bulge stars 
(Fabrizio et al. 2016). Moreover, the spectral energy distribution ($ugri$ bands) to separate 
Disk and Bulge stars (Calamida et al. 2017). The reddest ($zy$) LSST bands can overcome thorny 
problems with differential reddening (right panel in Fig.~2).

\vspace{-0.25truecm}
\begin{center}
\bf Why a deep minisurvey
\end{center}
\vspace{-0.35truecm}

$\bullet$ {\em Deep into the darkness} -- The current optical photometric survey 
is strongly limited in the two innermost degrees
above and below the Galactic plane. The absorption in these regions ranges from 
$A_K \sim 1$ to $A_K \sim 1.8$ mag, this means that in the visual band, 
$A_V$ ranges from 10 to almost 19(!) magnitudes. However, it is significantly smaller in
the redder LSST bands (see Fig.~3). Note that VVV is opening the path 
(Contreras Ramos et al. 2018), but the identification of variable stars is 
more difficult because the luminosity amplitude in the $K$-band is a factor of two 
smaller than in the $iz$-bands. This means that LSST can provide a complete
census of RRLs even in these highly reddened regions. 
These new reddening maps and MDFs cover the entire Bulge and the Galactic center, 
thus providing the opportunity to determine the density profile of old stellar 
populations. Note that the Bulge and the Halo density profiles in the inner 
regions of the Galaxy are expected to be different: the former being 
steeper than the latter (Wegg \& Gerhard 2013; P\'erez-Villegas et al. 2017; 
Kunder et al. 2018; Valenti et al. 2018). There is evidence that the Bulge includes a modest fraction of 
dark matter (15-20\%). This means a core or a mild cusp in the density profile of 
the dark matter halo (Portail et al. 2017) that can be easily traced with RRLs.

$\bullet$ {\em Stellar populations beyond the Galactic center} -- 
NIR time series data collected with 1-4m class telescopes have
revealed a sizable sample of classical Cepheids located in and 
beyond the Galactic center (Matsunaga et al. 2011;
D\'ek\'any et al. 2015; Matsunaga et al. 2018; Inno et al. 2019). 
This means the opportunity to investigate young (10-250 Myr) 
stellar tracers in a region of the Disk in which our knowledge of the radial 
distribution and its scale height is quite poor. More recently, 
Kains et al. (2018) found more than 2,500 variables in a modest 
FoV (VIMOS at VLT) and among them more than 100 are candidate CCs that 
appear to be located beyond the Galactic center. The limiting
magnitudes of the deep minisurvey will allow 
us to trace the young population in a large fraction of the Disk.

$\bullet$ {\em Absolute age distribution} --  Absolute age estimates based on the magnitude of the 
main sequence turn off (MSTO) are affected by uncertainties in distance and in reddening correction. The
difference in magnitude between the MS knee and the MSTO is independent of these uncertainties (Bono et al. 2010b).
This means absolute ages that are at least factor of two more accurate than the classical ones. We plan to trace 
the possible occurrence of multiple ancient star formation events using the MS knee in $izy$ bands (23.5-24.0 mag). 


{\bf Why LSST}. The Bulge is one of the main reasons why the ground-based observing
facilities are mainly developed in the Southern Hemisphere. The unique 
optical characteristics of LSST and the fact that it is the first 
experiment collecting deep multi-band time series data 
over a long time interval, will allow us to provide a complete census 
of the Bulge stellar content. 


\clearpage

\begin{figure}
\centering
\includegraphics[width=0.36\textwidth]{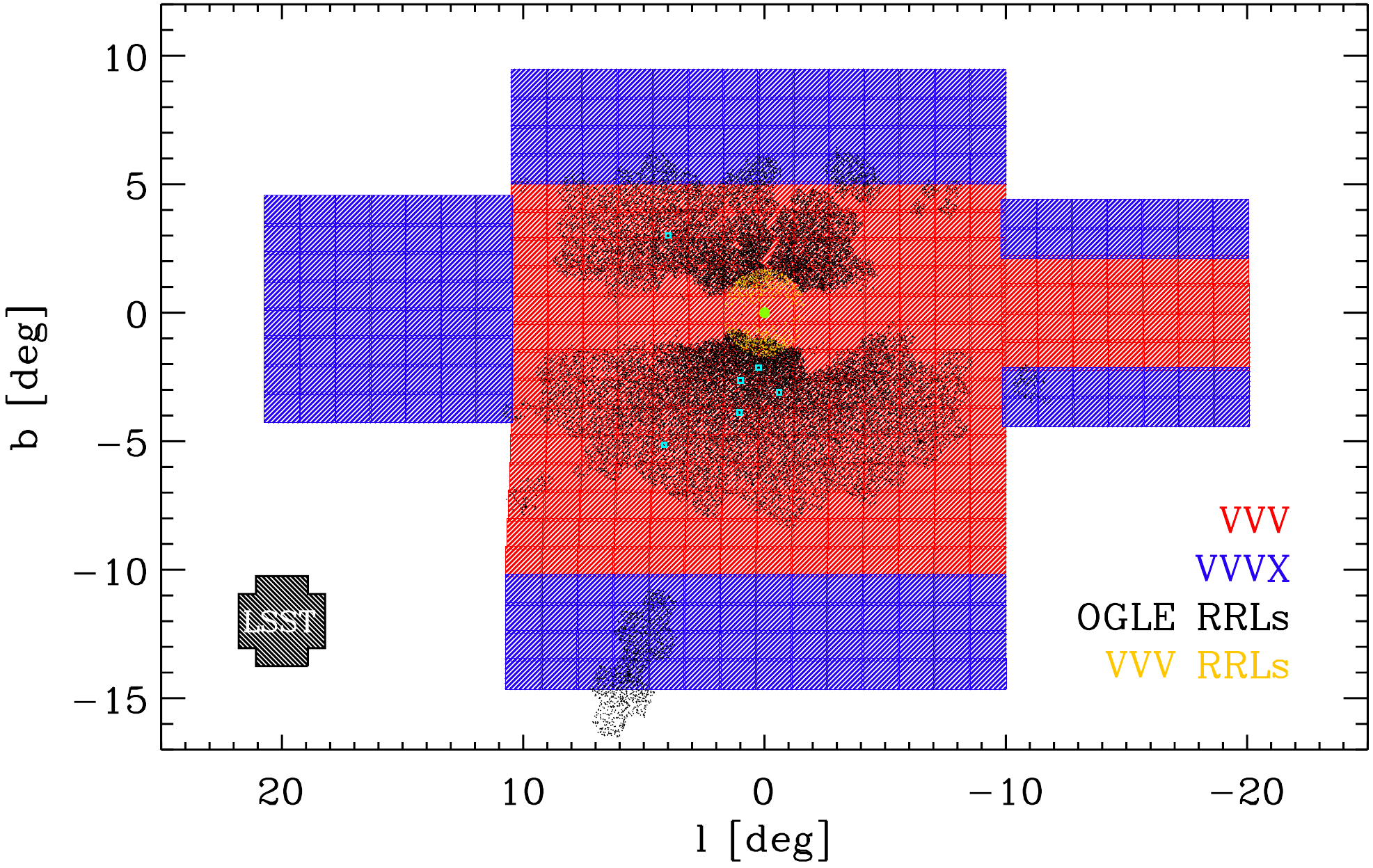}
\includegraphics[width=0.40\textwidth]{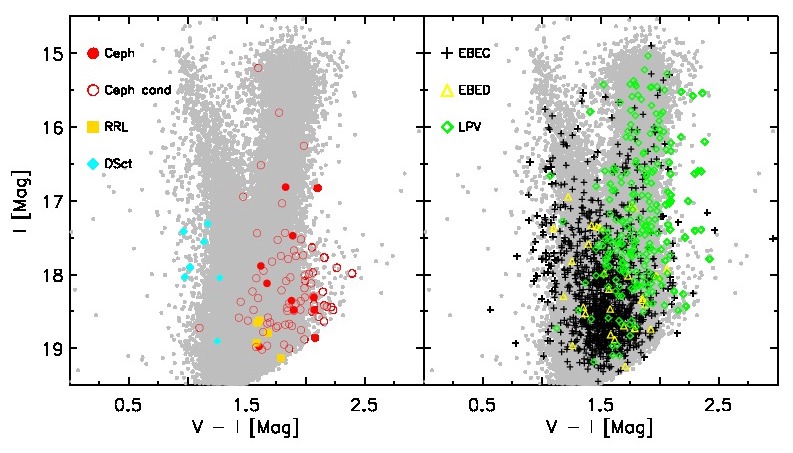}
\vspace{-0.5cm}
\caption{\footnotesize \textit{Left:}Distribution in Galactic coordinates of the Bulge 
and Thin Disk regions covered in the NIR bands by VVV and VVVX (red and blue boxes). 
The black dots display the RRLs detected by OGLE IV
(Pietrukowicz et al. 2015), while the yellow ones the RRLs detected by VVV (Contreras Ramos et al. 2018). The green
and the magenta circles mark the Galactic center and Bulge low-reddening regions (Dutra et al. 2002). The grey area
shows the FoV of LSST.
\textit{Right:} $I, V$ -- $I$ CMD of the new variables ($\sim$2,500) identified by Kains et al. (2018).}
\label{fig:fig1}
\end{figure}

\begin{figure}
\centering
\vspace{-0.3cm}
\includegraphics[width=0.78\textwidth]{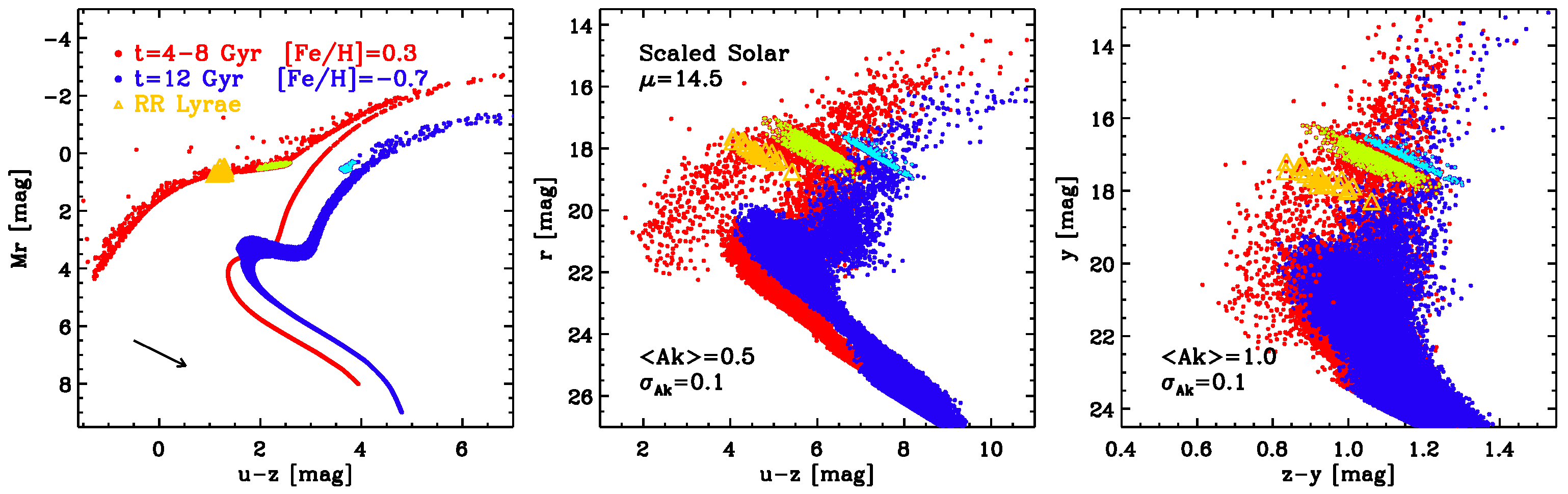}
\vspace{-0.3cm}
\caption{\footnotesize \textit{Left}: $r$,$u-z$ synthetic CMD for two different 
stellar populations characterized by different metal content and chemical composition 
(see labeled values). 
The yellow triangles display RRLs, light yellow dots mark red HB stars and old 
RGB bump, light blue dots RC and intermediate age RGB bump.  
\textit{Middle}: same as the left, but the stars were randomly perturbed by assuming a mean 
reddening typical of the Baade window ($A_K$=0.5 mag). 
\textit{Right}: $y$,$z-y$ CMD, with the stars were randomly perturbed from the theoretical CMD by 
assuming a mean reddening of $A_K$=1 mag.}
\label{fig:cmd}
\end{figure}

\begin{figure}
\vspace{-0.37cm}
\centering
\includegraphics[width=0.76\textwidth,trim={0cm 0.5cm 0cm 0cm},clip]{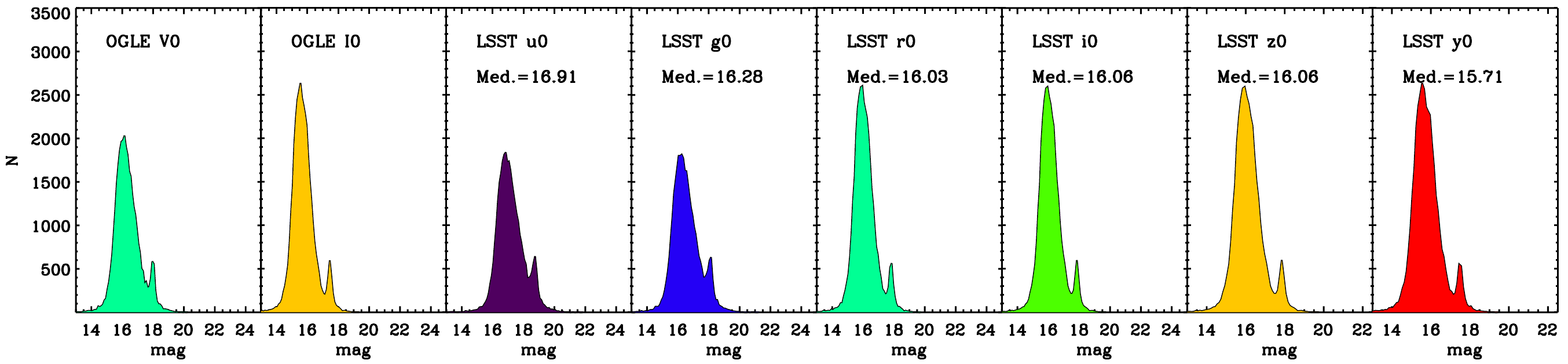}
\vspace{-0.2cm}
\includegraphics[width=0.76\textwidth]{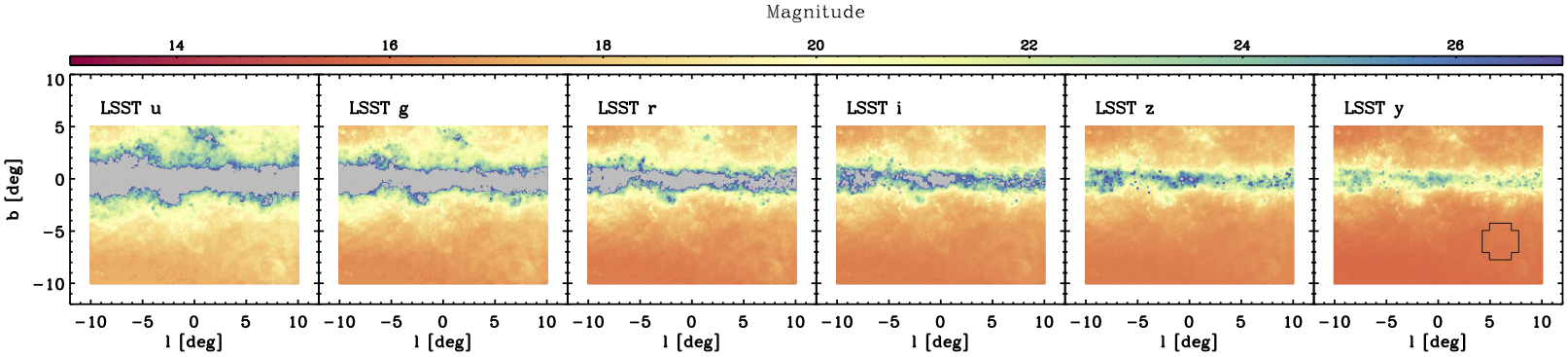}
\vspace{-0.3cm}
\caption{\footnotesize \textit{Top:} From left to right, un-reddened magnitude distributions of Bulge RRLs detected by OGLEIV ($V,
I$). They were un-reddened by using the reddening map provided by Gonzalez et al. (2012) reddening law by Cardelli
et al. (1989). The expected un-reddened magnitude distributions in LSST bands ($u, g, r, i, z, y$) are also displayed. The
$V,I$ bands were transformed into the LSST bands by using Jordi et al. (2006) and the mean RRL colors provided by
Vivas et al. (2017) and Coppola et al. (2011).
\textit{Bottom:} Apparent magnitude distribution of Bulge RRLs in 
Galactic coordinates for LSST bands. The color coding is plotted on top 
of the panels. The grey color marks areas in which the 
RRLs are fainter than 27 mag.}
\label{fig:distr_mag}
\end{figure}

\clearpage





\section{Technical Description}

\subsection{High-level description}

We plan to get shallow and deep exposures with the same cadence as the WFD survey, but alternating shallow and deep exposures (see below). We will then distinguish between a shallow minisurvey and a deep minisurvey. \\
It is important to stress that all the subsequent discussion is based on the experience of the TVS Crowded Field Photometry Task Force (CFTF), of which one of us is the chair. The goal of the CFTF was to study the efficiency of the detection of variable stars (mainly RRLs) in very crowded fields, and to study the best data analysis strategies to find and characterize the variables. We focused on a DECam dataset (NOAO 2013A-0719, PI: A. Saha) of the Bulge, with similar characteristics (photometric depth, crowdness, pixel scale) to LSST, and for which we already knew the variables form OGLE~4. The final outcome of the CFTF was that all the known bright variables, with our data analysis approach, were correctly retrieved. Furthermore, the identification of several new variables was made possible thanks to the use of a new period-search algorithm (Dall'Ora et al. 2019, in preparation).

\vspace{.3in}

\subsection{Footprint -- pointings, regions and/or constraints}
\begin{itemize}
\item The shallow minisurvey covers an area of -20$\lesssim$l$\lesssim$+20 deg and -15$\lesssim$b$\lesssim$+10 deg, in all the $u,g,r,i,z,y$ bands
\item The deep minisurvey is restricted to an area of -20$\lesssim$l$\lesssim$+20 deg and -3$\lesssim$b$\lesssim$+3 deg and to the $i,z,y$ bands.
\end{itemize}

\subsection{Image quality}
For the shallow minisurvey we can accept seeing of the order of $\sim 1$ arcsec. 
The deep minisurvey will be conducted in very highly crowded regions and, even if 
only the reddest bands (which provide smaller FWHMs) are requested, 
it is mandatory to ask for the median seeing at Cherro Pachon ($\sim 0.7$ arcsec).


\subsection{Individual image depth and/or sky brightness}

\begin{itemize}

\item The shallow minisurvey is made of 5s+5s exposures in all the bands. This will allow us to cover the magnitude ranges shown in Tab.~\ref{table_shallow}.

\item The deep minisurvey will be conducted only in the $i,z,y$ bands, aiming at identifing 
RRLs and RC stars in highly reddened regions of the Galactic Bulge. Indeed, according to 
Fig.~\ref{fig:distr_mag}, in the inner regions we expect to find the bulk of RRLs at $i,z \sim 24.5$ mag and at $y\sim$ 24 mag. To reach this limits with a SNR of at least $\sim 5$, we need exposure times of 60, 150 and 300s in the $i,z,y$ bands, respectively. Moreover, in the less reddened (i.e. more external) regions these limits will allow us to cover the faint end of the luminosity distribution of the RRLs.
\end{itemize}

The overall proposed strategy is to collect in the internal regions 
(see Footprint section) the shallow + deep exposures alternatively, 
according to the sequence: $u,g,r,(i_{shallow},i_{deep})$,
$(z_{shallow}$,$z_{deep}),(y_{shallow},y_{deep})$. 
This strategy minimizes the pointing and changing filter overheads. 
In the external Bulge regions we only plan to collect the shallow exposures.

\begin{table}
	\centering
	\caption{Expected saturation and $5\sigma$ limits for the shallow survey.}
	\label{table_shallow}
	\begin{tabular}{|ccc|}
		\hline
		band & saturation & $5\sigma$ \\
		     &  (mag)     &   (mag) \\
		\hline
		u & 13.5 & 22.2 \\
		g & 14.5 & 23.6 \\
		r & 14.6 & 23.1 \\
		i & 14.6 & 22.7 \\
		z & 14.1 & 22.1 \\
		y & 12.7 & 21.3 \\
		\hline
	\end{tabular}
\end{table}

These magnitudes have been computed with a custom ETC, which is based on the saturation and on the $5\sigma$ limits listed in https://smtn-002.lsst.io/ and in \\https://www.lsst.org/sites/default/files/docs/sciencebook/SB\_3.pdf.


We do not have special requirements on the sky brightness, since we ask for short exposures in the bluest bands. However, the SNR would benefit of grey time.


\subsection{Co-added image depth and/or total number of visits}

The final depth is not really relevant for this project, since we are interested in time-series of variable stars. However, we remark that the RC stars, which are static stars and that we use as population tracers, are already retrieved with a single visit. Finally, we note that the final stacked image will be of great interest for all the studies on the Galactic structure.


\subsection{Number of visits within a night}

There are no constraints on the number of visits per night, since we adopt the WFD cadence. 
However, since RRL light curves change significantly on short timescales, we ask a gap of 
at least one hour between two consecutive visits to the same pointing.
Moreover, we stress again that for the internal regions we ask to collect shallow and 
deep exposures one after the other to save the overhead time for the change of the filter.


\subsection{Distribution of visits over time}

There are no particular timings or scheduling requirements. We performed a MAF analysis with the PeriodicStarFit jupyter notebook, which adopts the WFD cadence, to check the efficiency of the adopted strategy (see sect. 4).


\subsection{Filter choice}
For the external, less reddened regions (shallow minisurvey) we ask for the full $u,g,r,i,z,y$ bands, since the full combination of all the filters allows us to use the REDIME technique and to disentangle the stellar populations of different metallicity. Moreover, they are also requested for the more internal regions, to study the foreground population. For the deep minisurvey, restricted to the internal regions only, we ask only for the $i,z,y$ bands, since they are less affected by the absorption. 

\subsection{Exposure constraints}
The 5s+5s exposures of the shallow minisurvey are designed to both avoid the saturation for the brightest RRL and RC stars, and to reach a reasonable depth at the $5\sigma$ level. Indeed, the median value of the magnitude distribution in all the bands is always reached with the total 10s exposure. The exposures of the deep minisurvey are dictated by the need to reach the RRLs in the more reddened regions.
Finally, it is important to note that the deep exposures saturate at a brightness level which is well within the dynamical range of the shallow survey, being $17.9, 18.9, 19.0, 19.0, 18.5, 17.1$ mag in the $u,g,r,i,z,y$ bands, respectively.


\subsection{Other constraints}

The current experiment has a excellent overlap with the LSST minisurvey suggested by Gonzalez and collaborators. 
The reason is twofold: 
i) they are very much interested in tracing static stars across the Bulge and in particular in highly reddened regions.  
ii) they plan to complement multiband optical photometry collected with LSST with NIR photometry collected by VVV and VVVX, 
radial velocity measurements with multi-object spectrographs, and in particular, with new kinematical and dynamical 
models of the MW formation and evolution. The two groups have a significant fraction of their science in common.  

The current experiment has no overlap with the minisurvey suggested by Clementini \& Musella, 
focussed on a sample of MW dwarf satellites, since in the area we plan to cover there is only 
one dwarf galaxy (Sagittarius dSph). However, their observing strategy is significantly different 
than the current one, since they plan to use the same exposure time of the WFD survey (15/30 sec), 
We are suggesting shorter exposures for the shallow minisurvey and longer exposures for the deep 
minisurvey. Note that the current observing strategy will allow us to identify the RRLs 
belonging to the Sagittarius dSph and to the Sagittarius stream, since they are on average 
$\sim$2.5 magnitude fainter than the RRLs in the Bulge. The Sagittarius stream has already been 
traced in the Halo and in the Bulge, but we still lack solid constraints on the Sagittarius 
stream in the innermost Galactic regions.

The same applies to the Galactic plane surveys proposed by R. Street and collaborators, and by M. Lund and collaborators. 
They propose to increase the cadence, in order to improve the detection of variable sources for 
a variety of science cases. However, the quoted WPs rely on the 15/30 sec WFD observing strategy, 
which affects the identification and characterization of both bright and faint/reddened stellar tracers 
we plan to use for our science. It is worth mentioning that the standard 15/30 sec visits
will allow us to trace some of the variables we are interested in, but their spatial distribution would 
be limited to partially reddened Bulge regions. This spotted sampling is far from being complete and 
homogeneous as required in a detailed survey. Note that the science drivers (REDIME) of the shallow 
minisurvey relies on the six $u,g,r,i,z,y$ LSST bands.

Note that Gaia will provide a complete census of both TIICs and Miras in low 
reddening regions of the Galactic Bulge. These objects will be saturated in almost all the LSST bands 
at the distance of the Bulge, thus further supporting the complementarity between the two experiments.

Moreover, we can determine proper motions for all of our sample, either over the 10 year LSST mission, or by using existing optical and near-infrared surveys as a first epoch.  Radial velocities and stellar abundances can be obtained with multi-object spectrographs such as MOONS/VLT, 4MOST/VISTA and AAOMEGA.

Finally, let us mention two independent astrophysical fields that will benefit a lot by the observing 
strategy and cadence we are proposing for both the shallow and the deep minisurvey. 

i) {\em Microlensing} -- The coupling between large FoV, cadence and number of visits in different 
photometric bands will provide unique opportunities to identify both short and long microlensing 
events (Navarro et al. 2018).  

ii) {\em Galactic Supernovae} -- There are reasons to believe that a significant fraction of Galactic 
supernovae are hidden by the Disk and the Bulge. The duration of the LSST experiment and the cadence 
we are suggesting will allow us to possible identify this rare event(s).  


\subsection{Estimated time requirement}

According to the LSST overheads, the expected total time for the internal fields (shallow + deep minisurvey) are:

\begin{itemize}
\item slew and setting the $u$ filter: 120s
\item 10 seconds (shallow) + 2 seconds shutter open/close: 12s
\item change filter to $g$ band: 120s
\item repeat in the $g$ band: 12s
\item change filter to $r$ band: 120s
\item repeat in the $r$ band: 12s
\item change filter to $i$ band: 120s
\item 10s (shallow) + 60s (deep) + 4s shutter: 74s
\item change filter to $z$ band: 120s
\item 10s (shallow) + 150s (deep) + 4s shutter: 164s
\item change filter to $y$ band: 120s
\item 10s (shallow) + 300s (deep) + 4s shutter: 314s
\end{itemize}

for a time requirement per visit of 1308 seconds (21.8 minutes).
Taking into account 825 visits, the total time per pointing is $\sim 299.75$ hours.\\

For the external regions we have:
\begin{itemize}
\item slew and setting the $u$ filter: 120s
\item 10 seconds (shallow) + 2 seconds shutter open/close: 12s
\item change filter to $g$ band: 120s
\item repeat in the $g$ band: 12s
\item change filter to $r$ band: 120s
\item repeat in the $r$ band: 12s
\item change filter to $i$ band: 120s
\item repeat in the $i$ band: 12s
\item change filter to $z$ band: 120s
\item repeat in the $z$ band: 12s
\item change filter to $y$ band: 120s
\item repeat in the $y$ band: 12s
\end{itemize}

for a time requirement per visit of 792 seconds (13.2 minutes).
Taking into account 825 visits, the total time per pointing is $\sim 181.5$ hours.\\

Since the area surveyed in the internal regions is $240$ square degrees ($25$ LSST pointings), the total
time requested for the deep minisurvey is $7,494$ hours.

The area surveyed in the external regions is $760$ square degress ($80$ LSST pointings), the total time
requested for the shallow minisurvey is $14,520$ hours.\\

In total, the time needed for both surveys would be $22,004$ hours. This number has to be compared to our estimated
time needed by the WFD survey to cover the same total area: $21,945$ hours ($912$ seconds $\times 825$ visits
$\times 105$ pointings).


\vspace{.3in}

\begin{table}[ht]
    \centering
    \begin{tabular}{|l|l|l|l}
        \hline
        Properties & Importance \hspace{.3in} \\
        \hline
        Image quality &    2 \\
        Sky brightness &  2 \\
        Individual image depth &  1 \\
        Co-added image depth & 3  \\
        Number of exposures in a visit   & 1  \\
        Number of visits (in a night)  & 2  \\ 
        Total number of visits & 2  \\
        Time between visits (in a night) & 2 \\
        Time between visits (between nights)  & 2  \\
        Long-term gaps between visits & 2 \\
        \hline
    \end{tabular}
    \caption{{\bf Constraint Rankings:} Summary of the relative importance of various survey strategy constraints.  1=very important, 2=somewhat important, 3=not important.}
        \label{tab:obs_constraints}
\end{table}

\subsection{Technical trades}

This is a long-term, time-series project. It will of course benefit from a good sampling of the light curves with as much as possible uniform image quality (FWHM, sky brightness). However, in our experience excellent results can be achieved even with highly non-uniform datasets [i.e. different instruments/telescopes with very different photometric depths and image quality, and non ad-hoc observing strategy, see e.g. Fiorentino et al. 2017]. Therefore, there are no really unfair trades for this project. The only constraint is the good seeing conditions for the inner regions of the Bulge, together with the proposed exposure times.


\section{Performance Evaluation}
The VESTALE observing strategy will allow us to secure 
accurate photometry  (1\% level) for both old (t~$>$ 10 Gyr) and 
intermediate age  (1$\lesssim$ t $\lesssim$ 9 Gyr)  stellar tracers. 
The LSST multi-band photometry will be compared with similar Bulge data  
collected with DECam at 4m Blanco telescope. This means the opportunity 
to validate the approach adopted to perform the photometry in crowded 
stellar fields, and in particular, the algorithms adopted to identify 
and to characterize stellar variability in crowded stellar fields 
(see the Task Force CFTF). Indeed, VESTALE overlaps with optical ($V, I$; OGLE IV, Kains et al. 2018, see Fig. 1, right panel), NIR ($Z, Y, J, H, K$; VVV, VVVX)
and SDSS ($u, g, r, i, z$, Vivas et al. 2017) photometric time series data.

To have a quantitative reference of the exected performance, we run a MAF simulation with the PeriodicStarFit jupyter notebook, that estimates the detected fraction of the input variable stars. 
By putting the RRLs expected reddened median (with respect to the total distribution) magnitudes, according to the simulation we can correctly retrieve $\sim 40\%$ of the periods after only one year (see Fig. 4). We stress that the plot shows the fraction of the detected variables at the median magnitude level, which is  $19.9,18.6,18.2,17.8,17.4,16.8$ mag in the $u,g,r,i,z,y$ bands, respectively. 
This means that the efficiency is higher at brighter magnitudes. Indeed, it is almost $100 \%$ after one year for the bright end of the magnitude distribution. Fig. 5 and Fig. 6 show the same analysis, but for the TIICs and the CCs. All the simulations are based on the known sample of variables, released by OGLE IV. In particular, we want to stress that the simulation on CCs is depicting the worst case (low period, small amplitude variables), and has to be considered as an ''acid test''.
It is worth mentioning that, after the first observing season, we are going to get efficiencies comparable to those of Gaia and of the currently available Galactic plane surveys.

As a technical comment, we stress that we could not change the (15 + 15) sec WFD visit in our simulations, and we simpy scaled, in the PeriodicStarFit jupyter notebook, the reference magnitudes by the differences in the expected flux ratios on the basis of our exposure times. Moreover, we ran our simulation over the entire WFD area. Our analysis could be improved with an {\textit {ad-hoc}} simulation on the actual area and with the actual exposure times. The PeriodicStarFit notebook is available in the standard \\
maf\_local/sims\_maf\_contrib/science/periodicVariables directory. It is based on the \\
/home/docmaf/maf\_local/sims\_maf\_contrib/mafContrib/periodicStarMetric.py code.

\begin{figure}
\centering
\includegraphics[width=0.5\textwidth]{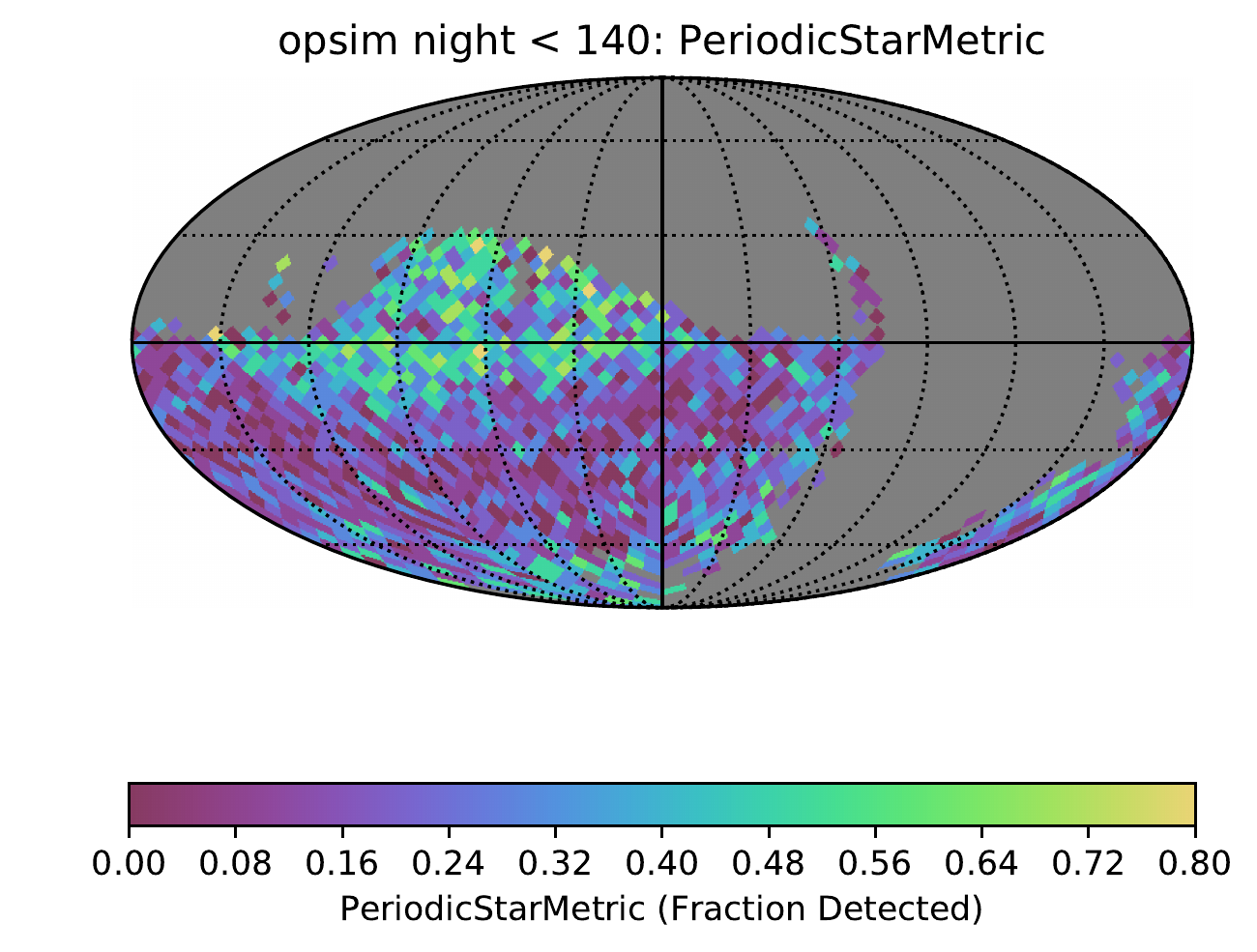}
\caption{\footnotesize Fraction of RR Lyrae stars detected after one year. For the simulation, we adopted an average period of 0.6d and an average amplitude of 0.5 mag at the expected median level of the RRLs magnitudes distribution. The map is based on the baseline2018a simulation, and it includes all the sky covered by the WFD survey. However, only the central part of the map is relevant for our project.}
\label{fig:opsim_140}
\end{figure}

\begin{figure}
\centering
\includegraphics[width=0.5\textwidth]{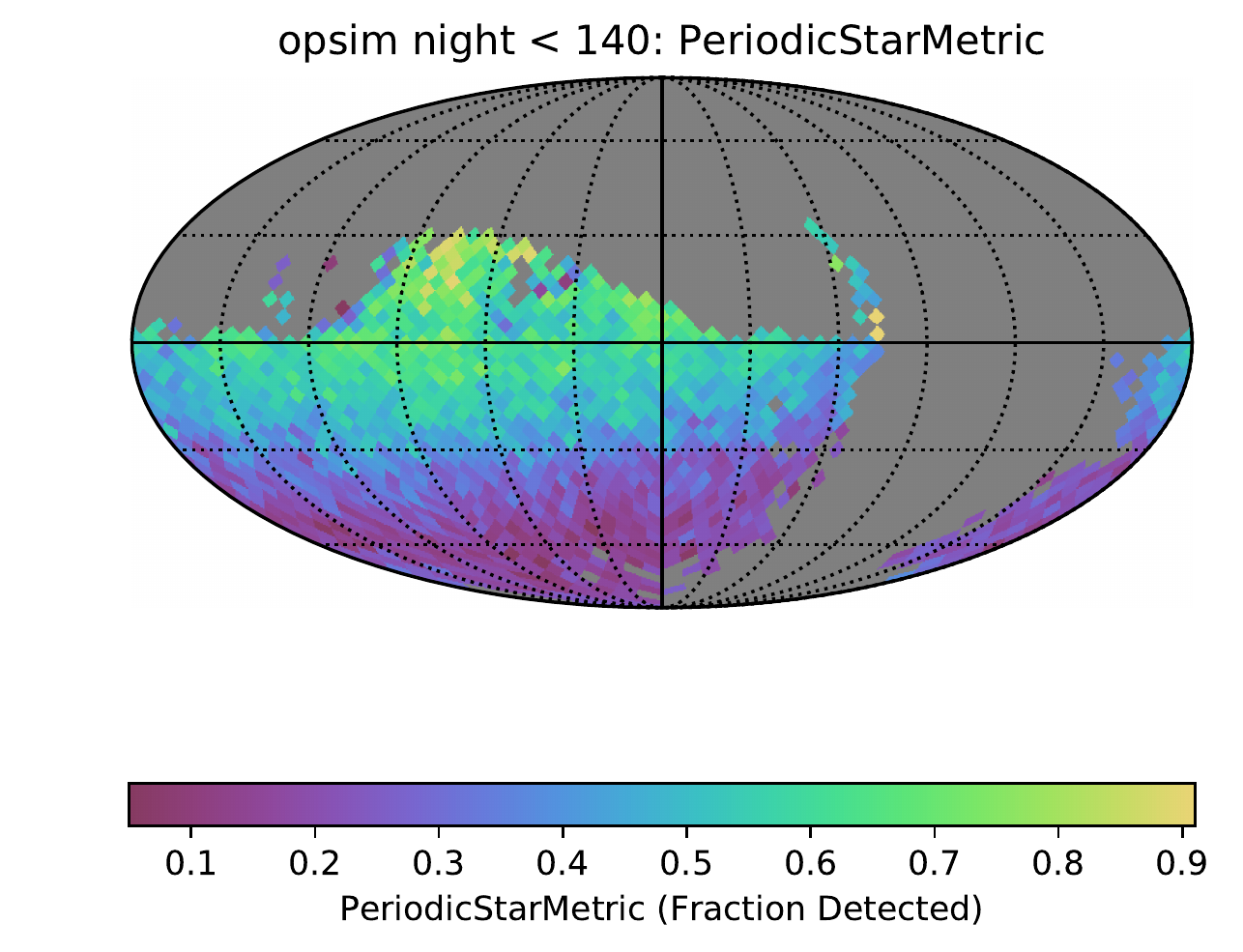}
\caption{\footnotesize Fraction of Type II Cepheids stars detected after one year. For the simulation, we adopted an average period of 2.0d and an average amplitude of 0.6 mag and at the expected median level of the TIICs magnitudes distribution. The map is based on the baseline2018a simulation, and it includes all the sky covered by the WFD survey. However, only the central part of the map is relevant for our project.}
\end{figure}

\begin{figure}
\centering
\includegraphics[width=0.5\textwidth]{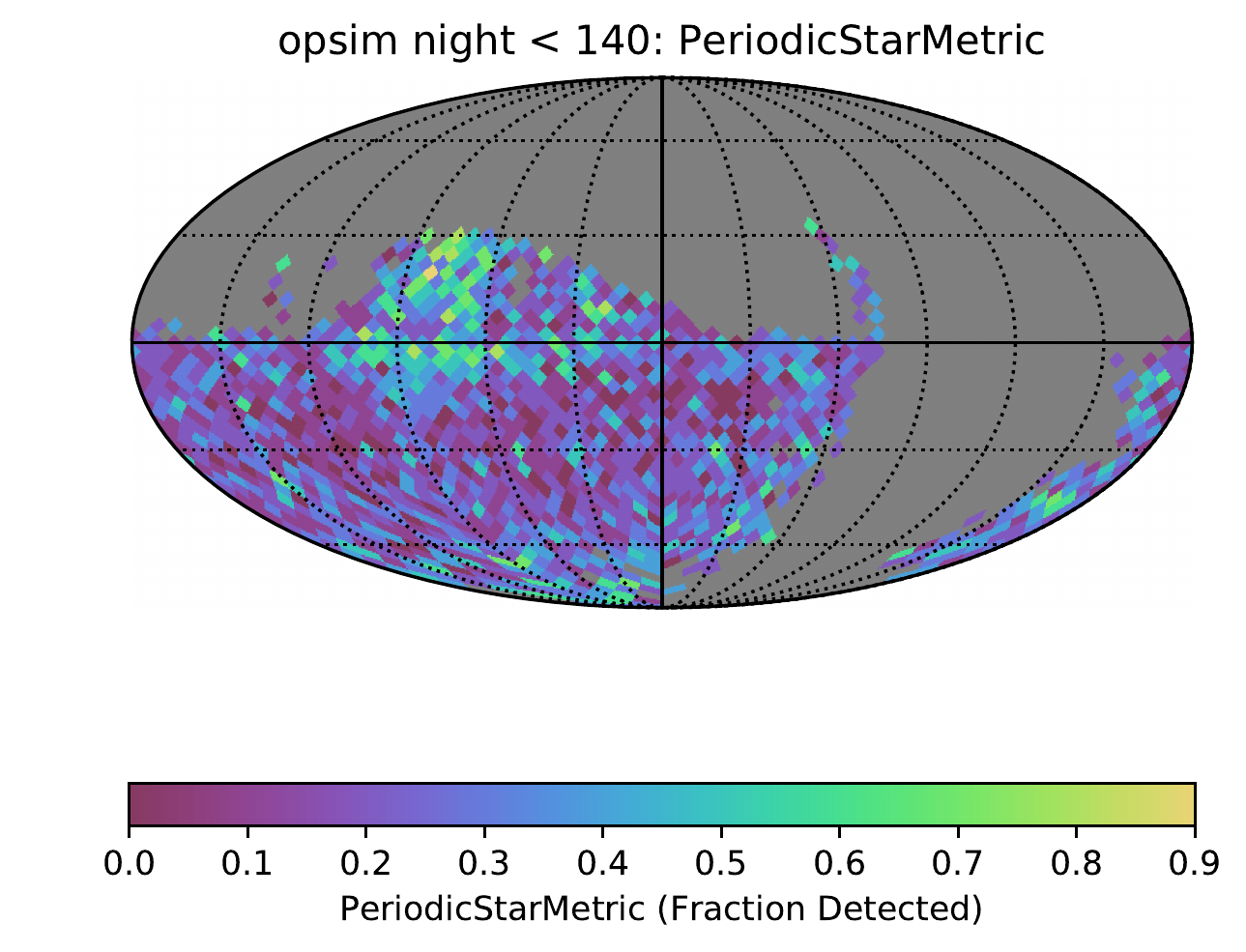}
\caption{\footnotesize Fraction of Classical Cepheids detected after one year. For the simulation, we adopted a period of 1.0d, corresponding to the mode of the CCs periods distribution, and an amplitude of 0.15 mag (corresponding to the typical amplitude of the 1.0d CCs) and at the expected median level of the 1.0d CCs magnitudes distribution. The map si based on the baseline2018a simulation, and it includes all the sky covered by the WFD survey. However, only the central part of the map is relevant for our project. Note also that in this simulation we studied the worst case, i.e. low period and small amplitudes CCs.}
\end{figure}


\vspace{.6in}

\section{Special Data Processing}

There are no special data processing requirements, since we adopt the same visit strategy as the WFD survey, with different exposure times.

\vspace{.6in}

\section{Acknowledgement} This work was developed within the Transient and Variable Stars Science Collaboration (TVS)
and the authors acknowledge the support of TVS in the preparation of this paper.

\section{References}
Athanassoula, E. 2005, MNRAS, 358, 1477 \\
Barbuy, B., Chiappini, C., \& Gerhard, O. 2018,ARA\&A, 56, 223 \\
Bono, G., Iannicola, G., Braga, V. F., et al. 2018, ApJ, accepted, arXiv181107069B \\
Bono, G., Stetson, P. B., Walker, A. R., et al. 2010a, PASP, 122, 651 \\
Bono, G., Stetson, P. B., VandenBerg, D. A., et al. 2010b, ApJL, 708, L74 \\
Bono, G., Genovali, K., Lemasle, B., et al. 2015, in ASPCS, Vol. 491, Fifty Years of Wide Field Studies in the Southern Hemisphere: Resolved Stellar Populations of the Galactic Bulge and Magellanic Clouds, ed. S. Points \& A. Kunder, 148 \\
Calamida, A., Strampelli, G., Rest, A., et al. 2017, AJ, 153, 175 \\
Cardelli, J. A., Clayton, G. C., \& Mathis, J. S. 1989, ApJ, 345, 245 \\
Carollo, D., Beers, T. C., Lee, Y. S., et al. 2007, Nature, 450, 1020 \\
Catchpole, R. M., Whitelock, P. A., Feast, M. W., et al. 2016, MNRAS, 455, 2216 \\
Contreras Ramos, R., Minniti, D., Gran, F., et al. 2018, ApJ, 863, 79 \\
Coppola, G., Dall’Ora, M., Ripepi, V., et al. 2011, MNRAS, 416, 1056 \\
D\'ek\'any, I., Minniti, D., Catelan, M., et al. 2013, ApJL, 776, L19 \\
D\'ek\'any, I., Minniti, D., Majaess, D., et al. 2015, ApJL, 812, L29 \\
Dutra, C. M., Santiago, B. X., \& Bica, E. 2002, A\&A, 381, 219 \\
Fabrizio, M., Bono, G., Nonino, M., et al. 2016, ApJ, 830, 126 \\
Fiorentino, G., Bono, G., Monelli, M., et al. 2015, ApJL, 798, L12 \\
Fiorentino, G., Monelli, M., Stetson, P. B., et al. 2017, A\&A, 599, A125 \\
Freeman, K., Ness, M., Wylie-de-Boer, E., et al. 2013, MNRAS, 428, 3660 \\
Gonzalez, O. A., Rejkuba, M., Zoccali, M., et al. 2012, A\&A, 543, A13 \\
Hammersley, P. L., Garz\'on, F., Mahoney, T. J., L\'opez-Corredoira, M., \& Torres, M. A. P. 2000, MNRAS, 317, L45 \\
Hill, V., Lecureur, A., G\'omez, A., et al. 2011, A\&A, 534, A80 \\
Inno, L., Urbaneja, M. A., Matsunaga, N., et al. 2019, MNRAS, 482, 83 \\
Jordi, K., Grebel, E. K., \& Ammon, K. 2006, A\&A, 460, 339 \\
Kains, N., Calamida, A., Rejkuba, M., et al. 2018, MNRAS, arXiv:1805.01898 \\
Kinman, T. D., Cacciari, C., Bragaglia, A., Smart, R., \& Spagna, A. 2012, MNRAS, 422, 2116 \\
Kunder, A., \& Chaboyer, B. 2008, AJ, 136, 2441 \\
Kunder, A., Rich, R. M., Koch, A., et al. 2016, ApJL, 821, L25 \\
Kunder, A., Valenti, E., Dall'Ora, M., et al. 2018, SSRv, 214, 90 \\
Matsunaga, N., Bono, G., Chen, X., et al. 2018, SSRv, 214, 74 \\
Matsunaga, N., Kawadu, T., Nishiyama, S., et al. 2011, Nature, 477, 188 \\
Matsunaga, N., Feast, M. W., Bono, G., et al. 2016, MNRAS, 462, 414 \\
McCarthy, I. G., Font, A. S., Crain, R. A., et al. 2012, MNRAS, 420, 2245 \\
Monelli, M., Milone, A. P., Fabrizio, M., et al. 2014, ApJ, 796, 90 \\
Navarro, M. G., Minniti, D., Contreras-Ramos, R., 2018, ApJ, 865, 5 \\
Ness, M., Freeman, K., Athanassoula, E., et al. 2012, ApJ, 756, 22 \\
P\'erez-Villegas, A., Portail, M., \& Gerhard, O. 2017, MNRAS, 464, L80 \\
Pietrukowicz, P., Koz\l owski, S., Skowron, J., et al. 2015, ApJ, 811, 113 \\
Rojas-Arriagada, A., Recio-Blanco, A., Hill, V., et al. 2014, A\&A, 569, A103 \\
Salaris, M., Percival, S., \& Girardi, L. 2003, MNRAS, 345, 1030 \\
Sch\"onrich, R., Asplund, M., \& Casagrande, L. 2014, ApJ, 786, 7 \\
Shen, J., Rich, R. M., Kormendy, J., et al. 2010, ApJL, 720, L72 \\
Stetson, P. B., Monelli, M., Fabrizio, M., et al. 2011, The Messenger, 144, 32 \\
Valenti, E., Zoccali, M., Gonzalez, O. A., et al. 2016, A\&A, 587, L6 \\
Valenti, E., Zoccali, M., Mucciarelli, A., et al. 2018, A\&A, 616, A83 \\
Vivas, A. K., Saha, A., Olsen, K., et al. 2017, AJ, 154, 85 \\
Wegg, C., \& Gerhard, O. 2013, MNRAS, 435, 1874  \\
Zoccali, M., Valenti, E., \& Gonzalez, O. A. 2018, A\&A, 618, A147 \\
Zoccali, M., Gonzalez, O. A., Vasquez, S., et al. 2014, A\&A, 562, A66 \\
Zoccali, M., Vasquez, S., Gonzalez, O. A., et al. 2017, A\&A, 599, A12 \\


\end{document}